\documentclass[12pt]{iopart}
\usepackage{graphicx,epsfig,subfigure,color,amssymb}

\begin{document}

\title{Imbibition through an array of triangular posts}
\author{M L Blow$^{1}$, H Kusumaatmaja$^{2}$, J M Yeomans$^{1}$}
\address{$^1$Rudolf Peierls Centre for Theoretical Physics, 1 Keble Road, Oxford, OX1 3NP,   
England}
\address{$^2$Max Planck Institute of Colloids and Interfaces,
am M\"{u}hlenberg 1, D-14476 Golm, Germany}
\ead{m.blow1@physics.ox.ac.uk}
\begin{abstract}
We present and interpret simulation results showing how a fluid moves on a hydrophilic   
substrate patterned by a square array of triangular posts. We demonstrate that the shape of   
the posts leads to anisotropic spreading, and discuss how this is influenced by the   
different ways in which the posts can pin the advancing front.  
\end{abstract}

\submitto{\JPCM}
\maketitle

\section{Introduction}
\label{introduction}
When a drop of liquid is placed on a surface it can either spread out to wet the surface, or   
form a spherical cap.
In the second, partial wetting, case the contact angle, between the tangent to the drop and   
the surface, is given by Young's equation \cite{young}
\begin{equation}
\cos\theta_{\mathrm{Y}}=\frac{\gamma_{\mathrm{SG}}-\gamma_{\mathrm{SL}}}{\gamma}
\end{equation}
where $\gamma_{\mathrm{SG}}$, $\gamma_{\mathrm{SL}}$ and ${\gamma}$ are the solid--vapour,   
solid--liquid and liquid--vapour surface tensions respectively.
Young's equation assumes that the contact line, where the liquid, vapour and surface meet,   
is able to move freely to enable the drop to relax to its equilibrium shape.

However, if there are surface heterogenieties, the contact line can be pinned leading to   
changes in the drop shape. Random impurites are difficult to treat theoretically, and cause   
problems in obtaining precise values for contact angles. However, as a result of recent   
advances in microfabrication techniques, it is becoming increasingly viable to fabricate   
surfaces which are micropatterned with posts or regular patches of different contact angle.   
These can be used to investigate the pinning behaviour of liquid drops and films, and have   
the potential to control fluids in microfluidic applications.

\begin{figure}
\centering
\epsfig{file=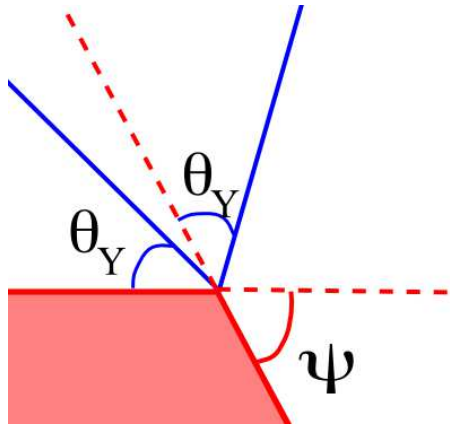,width=60mm}
\caption{Illustration of the Gibbs' Criterion. At a sharp corner, on a material surface, a   
wetting contact line remains pinned for all angles between the orientated equilibrium angles   
of the two planes: between $\theta_{\mathrm{Y}}$ and $\theta_{\mathrm{Y}}+\psi$}
\label{fig:gibbscriterion}
\end{figure}

Johnson and Dettre~\cite{johnsondettre} were the first to elucidate the nature and cause of   
contact line pinning. They placed drops axisymmetrically on substrates composed of   
concentric rings of hydrophilic and hydrophobic material and found that the contact line   
remained pinned on the hydrophilic-hydrophobic borders for contact angles between those of   
the two adjacent rings. The pinning occurs because both retreating to dewet the hydrophilic   
region, and advancing onto the hydrophobic region carry a free energy cost.
Oliver, Huh and Mason~\cite{huhmason,oliverhuhmason} recorded similar pinning at sharp   
corners. This is an example of the Gibbs' criterion which states that the contact angle will   
take a range of values spanning the dihedral angle of the corner (see   
Fig.\ref{fig:gibbscriterion}) ~\cite{gibbs}.
More recent theoretical and experimental work has studied how edge and corner pinning are   
important in controlling the shape of drops on surfaces patterned with chemical stripes or   
topographic ridges~\cite{khareetal,seemannetal,halimgrooves,chung}. The anisotropic pinning 
of the surface structure is reflected in the final shape of the drops. 
Moreover pinning on the tops of arrays of hydrophobic posts or ridges leads to   
superhydrophobic behaviour, associated with unusual drop hysteresis and dynamics~\cite{quere,reyssat}.

Capillary filling is also strongly affected by pinning effects caused by defects on the   
surface of a microchannel \cite{diotallevi,halimfilling}. Indeed, the filling is halted if   
there is a ridge lying across the channel, whose trailing edge makes an angle $\psi$ with   
the sides of the channel, if
\begin{equation}
\psi>180^{\circ} - 2\theta_{\mathrm{Y}}.
\end{equation}
This is due to the interface pinning on the corner of the ridge. An immediate   
consequence is that ridges with sides of differing slope can lead to anisotropic   
flow speeds, or unidirectional flow.

\begin{figure}
\centering
\epsfig{file=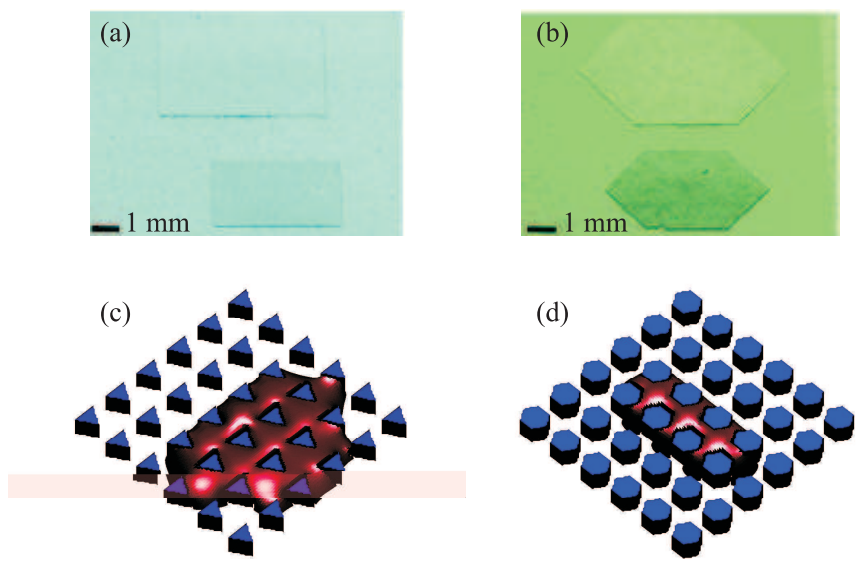,width=120mm}
\caption{Different drop shapes depending on (a-b) the arrangement and (c-d) the shape of
the posts. (a) Cylindrical posts which form a square lattice. (b) Cylindrical posts which
form
a hexagonal lattice. (c) Triangular posts which form a square lattice. (d) Hexagonal posts which form a   
square
lattice. The figures in panels (a) and (b) are experimental results taken from   
\cite{stone} (\textcopyright 2007 Nature Publishing Group), while
those in panels (c) and (d) are from our
simulations.}
\label{fig:fancyexamples}
\end{figure}

A situation simular to capillary filling is the spreading of liquid upon a single surface. Nonuniformities of the surface may lead to this spreading being aniostropic in nature, for example Cubaud and Fermigier~\cite{cubaud} noted that on a surface regularly patterned with chemical defects, a spreading drop forms faceted shapes, with 
the outer contact lines pinned on rows of defects. A related problem is the spreading of a thick film through an array of hydrophilic posts, prevented, again by pinning, from covering the tops of the posts, a   
phenomenon termed inbibition~\cite{bico,ishino2}. Courbin et al~\cite{stone} formulated   
pinning arguments describing the evolution of the fluid layer, together with a dynamical   
explanation of the final shape it would take. They argued that 
the top of the liquid film is pinned on the top edges of the posts, while the bottom of the   
film is able to move forward along the base. Spreading will occur if the film touches the   
next row of posts before it reaches the Young angle. The speed of this process depends on   
the post shape and spacing, and the Young angle. Therefore the drop can attain an   
anisotropic shape, with a symmetry reflecting that of the underlying lattice   
as shown in Fig.~\ref{fig:fancyexamples}. Courbin et al also noted that the advance of the imbitition front occurred via an `unzipping' transition as the contact line depinned consecutively from neighbouring posts, a process also identified by Sbragaglia et al~\cite{sbragaglia}.

Given the striking effect of post shape and anisotropy on capillary filling we might expect   
similar features to be observed in imbibition, an expectation confirmed by the results shown   
in Fig.~\ref{fig:fancyexamples}(c) and (d). The interplay between geometrical pinning and capillary forces may be extremely complex, making the shape of spreading drops difficult to predict. To investigate the different mechanisms that play a role, we focus in this paper on simulation   
results showing how a square array of trianglar posts is wet by an imbibing film as the   
contact angle is varied. The simulations allow us to follow the depinning of the interfaces   
in some detail, and relate it to the post geometry and the interface dynamics. We find that   
the spreading is indeed anisotropic, in a way that reflects the geometry of the triangles.   
The mechanism for depinning is strongly dependent on the details of the post geometry and on   
the Young angle, and depinning from both the top and the side edges of the posts is   
important in controlling the fluid behaviour.

%
%
\section{Simulation geometry}

\begin{figure}
\centering
\epsfig{file=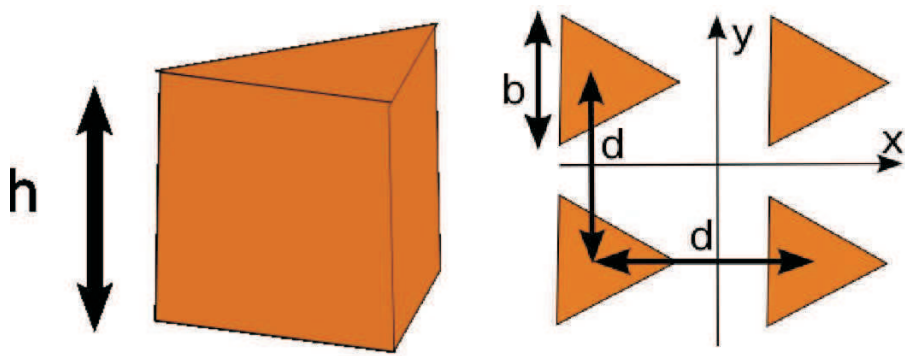,width=120mm}
\caption{Schematic diagram showing post dimensions.}
\label{fig:schematic}
\end{figure}
\begin{figure}
\epsfig{file=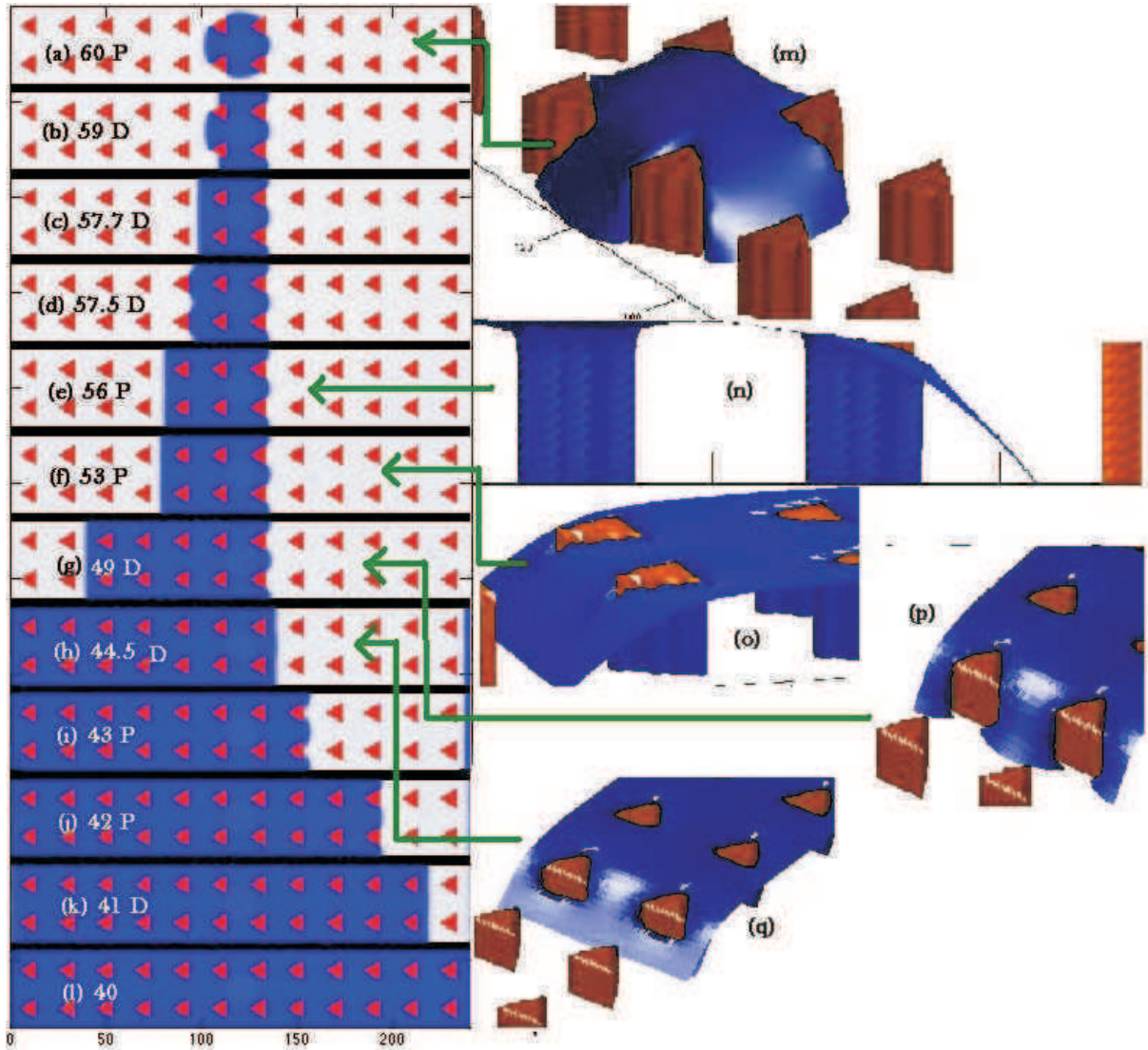,width=160mm}
\caption{(a)--(l): Plan view showing progressive wetting of the substrate (in blue) as the Young angle is slowly decreased and the water is introduced from a `virtual reserviour' at the centre of the sample. The Young angle is recorded on each view and P and D denote a pinned or dynamic state, respectively. (m)--(q): Three-dimensional views of the interface showing details of the pinning.}
\label{fig:multiplan}
\end{figure}
To understand the effect of pinning on the way a fluid spreads we consider, as a test case,   
imbibition through a square array of triangular posts. On grounds of symmetry, we would   
expect this geometry to show anisotropic spreading. We demonstrate that this is indeed the   
case, and indicate how the spreading depends on both the post shapes and the contact angle. 

The geometry we consider, illustrated in Fig.~\ref{fig:schematic}, is a square array, of   
lattice spacing $d$, of posts on a flat base. The cross-section of each post is uniform, and   
is an equilateral triangle, orientated to point along one of the primary axes of the array.   
We take this to be the positive $x$-direction and we will refer to it as the `forward'   
direction. The blunt ends of the posts face the `backward' direction, along $-x$. 

We denote the length of the sides of the posts by $b$, and their height by $h$. Thus the narrowest   
gap between triangles is along $y$ with spacing $d-b$, while the minimum spacing between the   
posts along $x$ is $d-\frac{\sqrt{3}}{2}b$. We consider equal post dimensions and a post   
spacing, $d=2h=2b$ (=20 lattice units). The posts and base substrate are taken to have the same, hydrophilic, Young angle.

We describe the equilibrium properties of the fluid in terms of a simple Landau free   
energy functional, and the dynamics by the Navier Stokes equations. This gives a diffuse   
interface model of the drop statics and dynamics, which we solve using a lattice   
Boltzmann algorithm. Details of the equations are given in the Appendix.

Imbibition is initiated by placing a small cylinder of liquid of radius 8 lattice units,   
extending between z=0 and z=h, at the centre of simulation box. The drop is replenished as it spreads by adding fluid from a   
virtual reservoir. This is done by setting  the density parameter at the starting location   
of the drop to the bulk equilibrium value in the liquid phase at each time step in the   
simulation. In this way, liquid is introduced while there is outwards flow, but once the   
interface is fully pinned, no new liquid enters the system.

\section{Imbibition in the $x$-direction}

To isolate the pinning effects in the $x$-direction we first consider a   
quasi-two-dimensional system, consisting of two rows of posts with periodic boundary   
conditions along $y$. We shall then compare similar simulations for spreading along $y$. We
slowly reduce the Young angle with the aim of observing the sequence of depinning   
transitions which occur. 
Simulations showing how a drop spreads along the $x$-axis, as the contact angle is reduced,   
are illustrated by the series of plan views in Fig.~\ref{fig:multiplan}. Three dimensional   
snapshots of the interface, at corresponding times, are also displayed. 
At the initial Young angle of $60^{\circ}$, the liquid is confined within its four closest   
posts, as shown in projection in Fig.~\ref{fig:multiplan}(a) and, in a full three   
dimensional view, in frame 4(m). The interface is pinned on the edges of the posts, unable to advance   
further. 

At a Young angle of $59^{\circ}$, the contact line inches forward across the surface to meet   
its periodic image along $y$. As shown in  Fig.~\ref{fig:multiplan}(b),(c), this enables the   
fluid to rearrange to form a configuration with the periodicity of the lattice along $y$. As   
it does this, fluid from both sides of the post tips meets. As a result the interface can no   
longer remain pinned to the tips and it depins to advance along $+x$. 

However, the interface almost immediately becomes pinned again: this time it is held back by   
the top of the blunt ends of the posts as shown in Figs.~\ref{fig:multiplan}(c), (n) and   
(o).  
On the base, the contact line has reached the Young angle and hence there is no capillary   
force driving the drop to spread. The mechanism behind this pinning is that any further   
progression of the fluid into the widening gap between the triangles, whilst maintaining the   
same angle with the substrate, leads to a thermodynamically unfavourable increase in   
interfacial area.

As the Young angle is decreased further the contact line is able to creep forward across the   
base, until it reaches the edges of the next posts. It quickly spreads forwards and upwards   
past the posts as shown in Fig.~\ref{fig:multiplan}(d), until it is pinned in the next gap   
Fig.~\ref{fig:multiplan}(e). The contact angle must be lowered further to overcome this pinning   
because the Laplace pressure, which drives the spreading, decreases as the drop size   
increases.  The interface finally reaches the next row of posts at a Young angle of about   
$51.5^{\circ}$.
Subsequent jumps become increasingly easy. This is because, as the drop gets bigger, the   
Laplace pressure varies less rapidly with size.

When the Young angle reaches $44.5^{\circ}$ the drop starts to move along $-x$. Three   
dimensional views of the drop configuration just before and just after this transition are   
shown in Figs.~\ref{fig:multiplan}(p),(q). At this Young angle, a   
portion of the interface near to the base is able to advance around the corners. Liquid from   
the two sides meet and the interface quickly moves upwards until the entire back faces of the posts   
are wet, thus allowing depinning.
As the interface now meets the substrate at an angle much larger than the Young angle, it advances   
rapidly towards the next set of posts. The flow is faster for these depinning events   
because they occur at lower contact angles and therefore, once depinning has occurred, the   
capillary force is stronger.

Once the liquid is moving towards $-x$, it does not advance in the $+x$ direction. This is a   
dynamical effect related to the way in which we refill the channel; the fluids is capable of   
moving in either direction, but does so more easily in the $-x$-direction. To isolate this effect we next consider filling in mirror-image geometries.

\subsection{Mirror-image substrates}
\begin{figure}
\centering
\epsfig{file=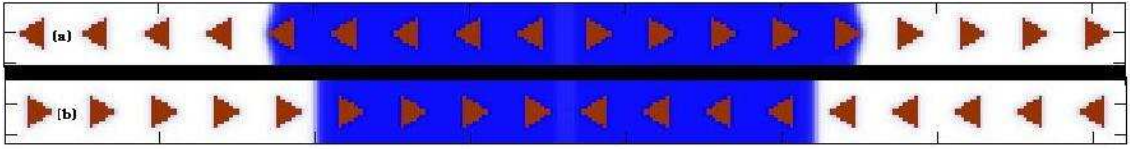,width=150mm}
\caption{Mirror substrate configuration. (a) forward facing (b) backward facing}
\label{fig:mirrorexample}
\end{figure}

\begin{figure}
\centering
\epsfig{file=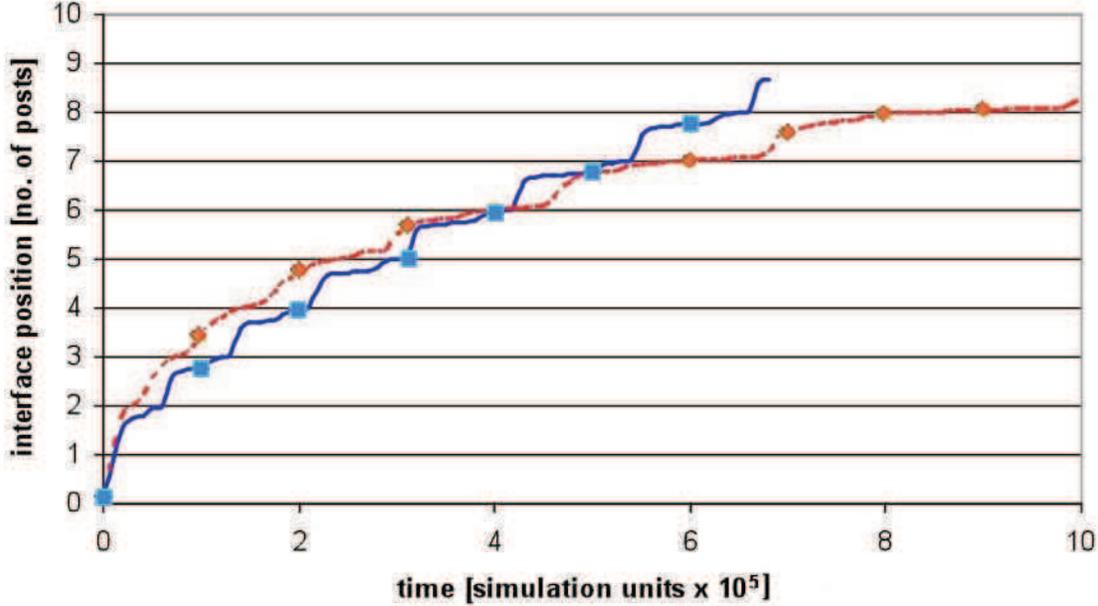,width=150mm}
\caption{Position of the contact line as a function of time for $\theta_{\mathrm{Y}}=40^{\circ}$ for the mirror-image substrates.
Dotted, red, diamonds: forward-facing triangles. Solid, blue, squares: backward-facing triangles. The integers on the y-axis lie at the centres of the gaps between posts.}
\label{fig:spreading}
\end{figure}

In our investigation of imbibition along $x$ we noted that flow in a given   
direction may be inhibited by either static effects (pinning on surface features) or dynamic   
effects (being out-competed by flow in the opposite direction). To isolate the former   
we performed simulations where the orientation of the triangular posts was invariant under reflection about the   
reservoir region (see Fig. \ref{fig:mirrorexample}). Now the pressure is the same in either   
direction and the contact line will only stop when a static   
equilibrium is reached.\\
\begin{figure}
\centering
\epsfig{file=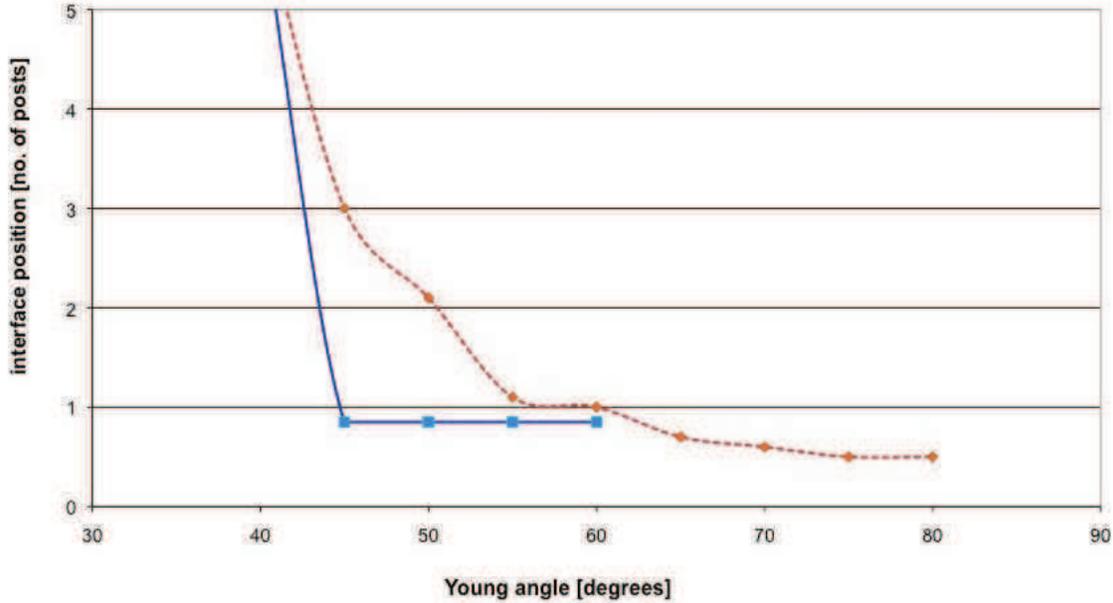,width=150mm}
\caption{The distance the contact line advances before pinning, as a function of Young angle. Dotted, red, diamonds: forward-facing triangles. Solid, blue, squares: backward-facing triangles. The integers on the y-axis lie at the centres of the gaps between posts.}
\label{fig:distanceforangle}
\end{figure}
We ran simulations for a range of constant Young angles and measured the final distance   
travelled by the contact line before pinning. Results, for the forward facing and backward facing triangles, are compared in Fig. \ref{fig:distanceforangle}. For $\theta_{\mathrm{Y}}\leq 40^{\circ}$ spreading continues indefinitely in both directions. When $\theta_{\mathrm{Y}}\geq 45^{\circ}$, there is pinning on the first row of posts for the backwards direction.   
There is also pinning in the forward direction, but not necessarily on the first posts; for $\theta_{\mathrm{Y}}= 45^{\circ}$ the interface advances   
past three posts, for $\theta_{\mathrm{Y}}=50^{\circ}$ two posts, and for $\theta_{\mathrm{Y}}=55^{\circ}$   
just one. For $\theta_{\mathrm{Y}}\geq60^{\circ}$ the interface is confined to the immediate vicinity of the   
reservoir.\\
These simulations also allow us to look more closely at the spreading dynamics. In   
Fig. \ref{fig:spreading} we plot the $x$-coordinate of the contact line (measured mid-way   
between the posts) as a function of time for $\theta_{\mathrm{Y}}=40^{\circ}$. We compare   
the two mirror-image systems, with posts pointing away from, and towards, the centre.
There is overall slowing of the spreading rate with distance from the source because the   
Laplace pressure is decreasing and because the viscous retardation increases with area of substrate wetted.
Superimposed upon this is an oscillation with the period of the patterning, which results   
from changes in the capillary force as the surface features are traversed. 
For both geometries the interface pins on the blunt end of the posts, and filling is most   
rapid when the contact line is between the posts. The sharper boost occurs when filling the   
gap in the backward direction. This is because the narrowing gap reduces interfacial area   
and hence promotes the advance of the interface.
\section{Imbibition in the $y$-direction}
\begin{figure}
\centering
\epsfig{file=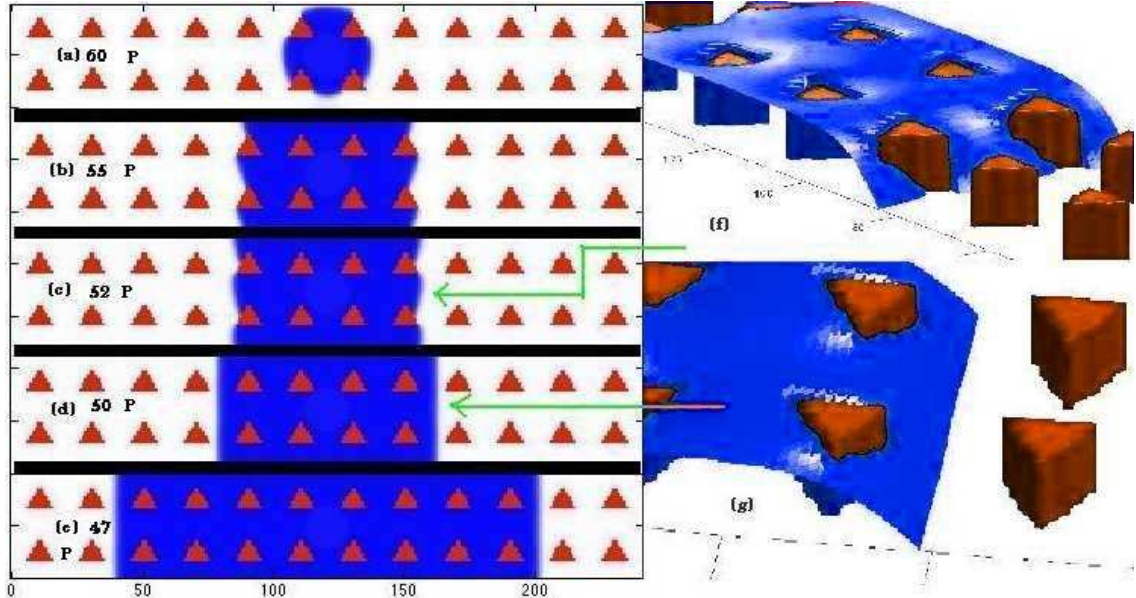,width=150mm}
\caption{(a)--(e): Plan view, showing (in blue) the regions of base substrate which are   
sequentially wet by the drop as the Young angle is decreased for filling in the   
$y$-direction. The Young angle, which is applied to both the base and posts, is recorded on   
each view. (f),(g) three dimensional views of the interface showing details of the   
pinning.
}
\label{fig:multiplan2}
\end{figure}

We now compare pinning for an interface spreading in the $y$-direction. We again consider a   
quasi-two-dimensional system, consisting of two rows of posts, but now with periodic   
boundary conditions along $x$, and follow the behaviour of the drop as the contact angle is   
decreased quasi-statically.
A series of plan views of the base of the drop at different times are shown in   
Fig.~\ref{fig:multiplan2}. Fig.~\ref{fig:multiplan2}(a) shows, as expected, a small drop   
pinned on the vertical corners of the four surrounding triangles. The interface joins with   
its periodic image along $x$ at $\theta_{\mathrm{Y}}\sim 56^{\circ}$.
Once this has occurred there is depinning along both $\pm y$, but the contact line stops at   
the subsequent rows of posts. 

Fig.~\ref{fig:multiplan2}(b) shows that the contact lines are pinned at the vertical edges   
of the triangles. With decreasing Young angle the contact line depins from the corner   
pointing along $+x$   
and creeps across the face of the triangle, as shown in Figs.~\ref{fig:multiplan2}(c),(f). 
Once it has crossed the side of the post it comes into contact with the interface in the   
adjacent gap and the contact line advances slightly.

However, it is almost immediately held up by pinning on the top of the posts, as shown in   
Figs.~\ref{fig:multiplan2}(c),(g). As for the $+x$ pinning discussed in the previous section, the contact line makes the Young angle with the base substrate, and is held from advancing further by the post. This pinning is weaker than that for motion along $x$ because the wall is at an angle to the   
advancing contact line. The contact line reaches the next row of posts at    
$\theta_Y=49^{\circ}$, and then spreads onwards, slowing at subsequent posts.

\section{Discussion}

The examples in Fig.~\ref{fig:fancyexamples} show that the arrangement and shape of posts on   
a hydrophilic surface can strongly affect the way in which a drop spreads on the surface. In   
this paper we have concentrated on the effect of post shape, and the way in which Gibb's   
pinning on post corners can affect the dynamics. As a simple system illustrating the   
behaviour we considered spreading on a square array of triangular posts.

A choice of two dimensional geometries allowed us to directly compare the dynamics in   
different directions relative to the orientation of the triangles. We found that the   
spreading is, as expected, anisotropic. Although depinning can occur at a higher contact   
angle in the forward direction (towards the points of the triangles), at lower contact   
angles the average flow velocity is higher in the backwards direction. This is because of a   
complex interplay between the pinning and the strength of the capillary forces which pull   
the drop through the array of posts.

Mechanisms of pinning are complex, and the interface can be pinned on the top or side edges   
of the posts. Indeed,
the importance of pinning phenomena in this geometry is highlighted by a comparison with the   
spreading condition for imbibition derived by Bico et al~\cite{bico} using averaged surface   
properties
\begin{equation}
\cos\theta_{\mathrm{Y}}\geq\frac{1-\phi}{r-\phi}
\end{equation}
where $r$ is the ratio of substrate area to projected area and $\phi$ is the ratio of the area of the post-tops to the projected area. For our geometry this criterion becomes $\theta_{\mathrm{Y}}\leq57.1^{\circ}$, which is a significantly higher threshold than our observed value of $40^{\circ}-45^{\circ}$.

In further work we will compare results for the full three dimensional   
geometry, where the connectivity of the interface will introduce new effects. Experiments on cylindrical posts have shown a range of final shapes for imbibition, depending upon surface parameters~\cite{stone}, and we aim to investigate how this picture is complicated by the shape of the posts and the consequent pinning. We hope that our calculations will stimulate further experiments on the effects of post shape on   
imbibition, and on the related problem of capillary filling. Understanding the interplay between   
pinning and capillary action more fully could help to design novel ways of controlling   
fluids in microfluidic channels.

\appendix
\section{Model details}
In the simulations reported here, the liquid drop and its surrounding gas are modelled using   
a one-component, two-phase fluid. The equilibrium properties of the system are described by   
a Landau free energy functional of the form
  \begin{equation} 
    \Psi = \int_V (\psi_b(n)+\frac{\kappa}{2} (\partial_{\alpha}n)^2) dV
    + \int_S \psi_s(n_s) dS
    \label{Aeq3}
    \end{equation}   
The first term $\psi_b(n)$ is a bulk free energy term which we take to be \cite{Briant}:
    \begin{equation}
    \psi_b (n) = p_c (\nu_n+1)^2 (\nu_n^2-2\nu_n+3-2\beta\tau_w)
    \end{equation}                                      
where $\nu_n = {(n-n_c)}/{n_c}$, $\tau_w = {(T_c-T)}/{T_c}$ and $n$, $n_c$, $T$, $T_c$ and   
$p_c$ are the local 
density, critical density, local temperature, critical temperature and critical pressure of   
the fluid  
respectively. The parameter $\beta$ is related to the density contrast between the liquid   
and gas phases.
This choice of free energy leads to two coexisting bulk phases (liquid and gas) of density 
$n_c(1\pm\sqrt{\beta\tau_w})$. 

The second term in Eq.\ (\ref{Aeq3}) models the free energy associated with any 
interfaces in the system. The parameter $\kappa$ is related to the surface tension through   
$\gamma =  
{(4\sqrt{2\kappa p_c} (\beta\tau_w)^{3/2} n_c)}/3$ and the interface thickness through $\xi   
= \sqrt{\kappa
n_c^2/4\beta\tau_wp_c}$ \cite{Briant}. 

The final term in Eq.\ (\ref{Aeq3}) describes the interactions between the 
fluid and the solid surface. Following Cahn \cite{Cahn}, the surface energy density is taken   
to be $\psi_s (n) =  
-\phi \, n_s$, where $n_s$ is the value of the fluid density at the surface. The strength of   
the interaction 
is parameterised by the variable $\phi$, and is related to the contact angle, $\theta_Y$ by   
\cite{Briant}
\begin{equation}
\phi = 2\beta\tau_w\sqrt{2p_c\kappa}
\mathrm{sign}(\frac{\pi}{2}-\theta_Y)\sqrt{\cos{\frac{\alpha}{3}}(1-\cos{\frac{\alpha}{3}})}   
\end{equation}
where $\alpha=\cos^{-1}{(\sin^2{\theta_Y})}$ and the function sign returns the sign of its   
argument. 
In the simulations, this equilibrium wetting boundary condition can be  implemented by setting   
the gradient 
of the density perpendicular to the solid surface to
\begin{equation}
\partial_{\perp}n = -\phi/\kappa
\end{equation}
The hydrodynamics of the drop is described by the continuity and the Navier-Stokes   
equations
\begin{eqnarray}
    &\partial_{t}n+\partial_{\alpha}(nu_{\alpha})=0
    \label{Aeq1}\\
    &\partial_{t}(nu_{\alpha})+\partial_{\beta}(nu_{\alpha}u_{\beta}) = 
    - \partial_{\beta}P_{\alpha\beta}+ \nu \partial_{\beta}[n(\partial_{\beta}u_{\alpha} +  
\partial_{\alpha}u_{\beta} 
+ \delta_{\alpha\beta} \partial_{\gamma} u_{\gamma}) ]
    \label{Aeq2}
\end{eqnarray}
where $\mathbf{u}$, $\mathbf{P}$, and $\nu$ are the local velocity, pressure tensor, and   
kinematic viscosity 
respectively. No-slip boundary conditions on the velocity are imposed on the surfaces   
adjacent to and opposite the drop and 
periodic boundary conditions are used in the two perpendicular directions. The thermodynamic   
properties of the system appear 
in the equations of motion through the pressure tensor $\bf{P}$; mechanical equilibrium is   
equivalent to minimising the free 
energy. $\bf{P}$ can be calculated from the free energy \cite{Briant} to give
\begin{eqnarray} 
&P_{\alpha\beta} = (p_{\mathrm{b}}-\frac{\kappa}{2} (\partial_{\alpha}n)^2 -
\kappa n \partial_{\gamma\gamma}n)\delta_{\alpha\beta} 
+ \kappa (\partial_{\alpha}n)(\partial_{\beta}n),  \\ 
&p_{\mathrm{b}} = p_c (\nu_n+1)^2 (3\nu_n^2-2\nu_n+1-2\beta\tau_w) .
\end{eqnarray}

Eqs.~(\ref{Aeq1}) and (\ref{Aeq2}) are solved using a Lattice Boltzmann algorithm.
Details of this approach, and of its application to
drop dynamics, can be found in \cite{Briant,Dupuis,Pooley,Yeomans}.

We have chosen the following parameters for the simulations: $\kappa = 0.01$, $p_c = 1/8$,   
$n_c = 3.5$, 
$T = 0.4$, $T_c = 4/7$, $\beta = 1$, $\tau_L = 2$ and $\tau_L = 0.7$. These correspond to an   
interfacial thickness 
$\xi = 0.9$, surface tension $\gamma = 0.029$, viscosity ratio ${\eta_L}/{\eta_G} = 7.5$,   
and density ratio ${n_L}/{n_G} = 3.42$  (all in lattice units).

\end{document}